\documentclass[aps,prl,twocolumn]{revtex4}
\begin{document}
\draft
\def\be{\begin{equation}}
\def\ee{\end{equation}}
\def\ba{\begin{eqnarray}} 
\def\ea{\end{eqnarray}}
\def\d{\mathrm{d}}

\newcommand{\bbf}{\mathbf}
\newcommand{\rrm}{\mathrm}

\title{ A Renormalisation Approach to Effective Interactions in Hilbert Space\\}
\author{Jean Richert\\
Laboratoire de Physique Th\'eorique, UMR 7085 CNRS/ULP,\\
Universit\'e Louis Pasteur, 67084 Strasbourg Cedex,\\ 
France} 
%
%
 
\date{\today}

\begin{abstract}
The low-lying bound states of a microscopic quantum many-body system of $n$
particles and the related physical observables can be worked out in a
truncated $n$--particle Hilbert space. We present here a non-perturbative
analysis of this problem which relies on a renormalisation concept and work out
the link with perturbative approaches.   
\end{abstract}
\maketitle


The ab initio construction of a rigorous quantum many-body theory able to 
describe bound particle systems like molecules, atoms, aggregates and atomic 
nuclei has developed over a long period of time starting in the sixties
~\cite{brand}. The quantities of interest are the 
spectroscopic properties of low-lying many-body states of the systems, their 
energies and  wavefunctions from which it is possible to obtain informations 
about relevant physical quantities like multipole moments, electromagnetic 
transition rates, spectroscopic factors and other experimentally accessible
observables.   
  
In practice the explicit resolution of the problem necessitates the 
diagonalisation of the
many-body Hamiltonian in Hilbert space which is spanned by a complete set of
basic states, in principle of infinite dimension, in any case generally very
large. The information of interest is however restricted to the  knowledge of
a few energetically low-lying states which often possess collective properties.
Hence it would be convenient to work in a finite truncated subspace of the 
original Hilbert space. Since the physical properties of the system should not
depend on the dimensionality of the space of states the original Hamiltonian 
has to be redefined
and gets an effective Hamiltonian which acts in the restricted space. The 
modification corresponds to a renormalisation of the $k$-body (generally 
$k$=1,2) interactions which govern the system.

Many different renormalisation procedures have been proposed and successfully
applied in the framework of the shell model and related microscopic 
descriptions. Rigorous projection methods lead to effective Hamiltonians which
can in principle be explicitly generated by means of perturbation techniques
~\cite{brand,kuo1,john,ober,sch1}.
Unfortunately there exists no straightforward control on the convergence
properties of the perturbation expansions, especially not when the interaction 
between the particles is strong as it is the case in atomic
nuclei for instance~\cite{bruce}. Many attempts have been made in order to 
overcome this problem ~\cite{suz,kum,hof}.
More pragmatic procedures have also been introduced such as two-body matrix 
elements corresponding to truncated perturbation expansion contributions whose 
use may be justified by means of physical arguments~\cite{ger}. Even more 
phenomenologically the interaction has been parametrised in terms of $2$-body
matrix elements
which are fitted in such a way as to reproduce an
arbitrarily selected number of experimental quantities~\cite{wil}. Very
recently effective $2$-body interactions have been constructed by means of a
non-perturbative renormalisation technique which cuts-off the large momentum
components of the interaction~\cite{bog}. Consequently this procedure eliminates
the effects of the strong repulsion at short distances which affect essentially
the high-excited states of the spectrum. 

All these methods are phenomenological and adapted to explicit calculations.
However they do not tackle the problem related to the breakdown of perturbation
methods in specific but frequently encountered situations. This is the case in 
the framework of the projection  method when two states belonging to different 
subspaces come arbitrarily close to each other~\cite{sch2}. 


The investigations which follow are developed in the spirit of former work
based on renormalization concepts ~\cite{gla,rau,mue,bec}. We introduce a 
non-perturbative
formulation of the bound state many-body problem. It grounds on the dimensional
reduction of Hilbert space which generates an effective Hamiltonian 
characterized by running coupling constants. The use of projection methods
allows to link the present approach to 
perturbative methods. We show explicitly the connection between
the divergence of perturbation expansions and the concept of fixed points. 

{\it Formal framework}.
Consider a
system with a fixed but arbitrary number of bound particles in a Hilbert space
$\cal H^{(N)}$ of dimension $N$ governed by a Hamiltonian
$H^{(N)}\left ( g_1^{(N)},g_2^{(N)},...g_p^{(N)}\right)$ where
$\left\{ g_1^{(N)},g_2^{(N)},\cdots, g_p^{(N)}\mapsto g^{(N)}\right\} $ 
are a set of 
parameters (coupling constants) which enter the expression of the interaction
operators in $H^{(N)}$. The eigenvectors $|\Psi_i^{(N)}(g^{(N)})\rangle$
$\left\{ i=1,\cdots, N\right\}$ span the Hilbert space and are the solutions of the
Schroedinger equation
\be
H^{(N)}(g^{(N)}) |\Psi_i^{(N)}(g^{(N)})\rangle = \lambda_i(g^{(N)}) 
|\Psi_i^{(N)}(g^{(N)})\rangle  
\label{eq0} \ .
\ee
The solutions are obtained by means of a diagonalisation which fixes both the
eigenvalues $\left\{\lambda_i(g^{(N)}) , \, i=1,\cdots, N\right\}$ 
and
eigenvectors $\left\{|\Psi_i^{(N)}(g^{(N)})\rangle,\, i=1,\cdots,N\right\}$ 
in terms of a linear combination of orthogonal basis states
$\left\{|\Phi_i\rangle, \,i=1,\cdots, N\right\}$.
Since dim ${\cal H}^{(N)} = N$ is generally very large if not infinite and the
information needed reduces to a finite part of the spectrum it makes sense to
try to restrict the space dimensions such that the physical quantities
related to  the part of the system which is of interest remain the same as
those obtained in the original space ${\cal H}^{(N)}$. If the relevant
quantities of interest are for instance $M$ eigenvalues out of the set
$\left\{\lambda_i(g^{(N)})\right\}$ then  
\be
H^{(M)}(g^{(M)}) |\Psi_i^{(M)}(g^{(M)})\rangle = \lambda_i(g^{(M)}) 
|\Psi_i^{(M)}(g^{(M)})\rangle  
\label{eq1} \ 
\ee
with the constraints
\be
\lambda_i(g^{(M)}) = \lambda_i(g^{(N)})
\label{eq2} \ 
\ee
for $i = 1,...,M$.
Eq. (3) implies relations between the sets of coupling constants $g^{(M)}$ and 
$g^{(N)}$ 
\be
g_k ^{(M)} = f_k(g_1^{(N)},g_2^{(N)},...g_p^{(N)})
\label{eq3} \ 
\ee
with $k = 1,...,p$.
The solution of this set of equations generates a new, effective Hamiltonian 
whose spectrum in the restricted space ${\cal H}^{(M)}$ is the same as the 
corresponding  one in the original complete space. 
 

{\it System at temperature T = 0}.
We now develop a general approach  which allows to follow the evolution of the
effective Hamiltonian of the system when the dimensions of the Hilbert space
are systematically reduced.   
Using the Feshbach formalism ~\cite{fesh,blo} we divide the Hilbert space
${\cal H}^{(N)}$ into two subspaces, $P{\cal H}^{(N)}$ and $Q{\cal H}^{(N)}$
with
\be
\mathrm{dim}\,  P{\cal H}^{(N)}= N-1, \quad \mathrm{dim}\, Q{\cal H}^{(N)}= 1
\label{eq10} \ .
\ee 
In the projected subspace $P{\cal H}^{(N)}$ the system with energy $E$ is 
described by the effective Hamiltonian
\be
H_{eff}(E) = P H P  +  P H Q {(E-Q H Q)}^{-1} Q H P
\label{eq11} \ . 
\ee
where we define
\be
H = H_0 + g H_1 
\label{eq12} \  
\ee
The set of basis states $\left\{|\Phi_i\rangle, \,i=1,\cdots, N\right\}$ are
the eigenvectors of $H_0$ with the corresponding eigenvalues 
$\left\{\epsilon_i, \,i=1,\cdots, N\right\}$.
The expression $H_{eff}(E)$ is generally the starting point of theories
which use perturbation expansions ~\cite{brand}. Here we proceed
differently. We consider 
\be
P |\Psi_1^{(N)}\rangle = \sum_{i=1}^{N-1} a_i^{(N)}(g^{(N)})|\Phi_i\rangle
\label{eq13} \  
\ee
which is the projection in $P{\cal H}^{(N)}$ of an eigenvector 
\be
|\Psi_1^{(N)}\rangle = \sum_{i=1}^{N} a_i^{(N)}(g^{(N)})|\Phi_i\rangle
\label{eq14} \  
\ee
of ${\cal H}^{(N)}$. If $\lambda_1^{(N)}$ is the eigenvalue
corresponding to  $|\Psi_1^{(N)}\rangle$ we look for the solution of 
\be
H_{eff}(\lambda_1^{(N)})P |\Psi_1^{(N)}\rangle =  \lambda_1^{(N)}
P |\Psi_1^{(N)}\rangle
\label{eq15} \ . 
\ee
We consider $P |\Psi_1^{(N)}\rangle$ to be the lowest eigenstate and 
$Q{\cal H}^{(N)}$
to contain f. i. the highest one. We impose the lowest eigenvalue in the 
$P{\cal H}^{(N)}$ subspace to be the same as the one in the complete space
\be
\lambda_1^{(N-1)} = \lambda_1^{(N)} 
\label{eq16} \ . 
\ee
Projecting this expression on $\langle\Phi_1|$ gives a relation  which fixes
$g^{(N-1)}$,  the coupling constant which characterises the strength of the 
interaction in the $P{\cal H}^{(N)}$ subspace in such a way that
Eq.~(\ref{eq16}) is verified
\be
 \langle\Phi_1|H_{eff}(\lambda_1^{(N)})|P |\Psi_1^{(N)}\rangle =\lambda_1^{(N)}
(g^{(N)})a_1^{(N)}(g^{(N)})
\label{eq17} \ . 
\ee
Then
\be
\langle\Phi_1|H_{eff}(\lambda_1^{(N)})|P \Psi_1^{(N)}\rangle = \cal
F(g^{(N-1)})
\label{eq18} \ . 
\ee
The r.h.s. of Eq.~(\ref{eq18}) can be worked out explicitly by using $H^{(N-1)}$
given by Eq.~(\ref{eq11}) in the expression of $H_{eff}(E = \lambda_1^{(N)})$.
The denominator in the second term of
$H_{eff}(\lambda_1^{(N)})$ is a scalar quantity since 
$\mathrm{dim}\, Q{\cal H}^{(N)}= 1$
where $N$ designates the highest lying energy eigenstate of $H^0$,  
$|\Phi_N\rangle$. One gets
\be
\cal F(g^{(N-1)}) = \overline{H_{1N}^{(N-1)}} + H_{1N}
(\lambda_1^{(N)} -  H_{NN})^{-1}\overline{H_{N1}^{(N-1)}}
\label{eq19} \ 
\ee
with 
\be
H_{ij} = \langle \Phi_i|H^{(N-1)}|\Phi_j\rangle 
\label{eq20} \  
\ee
and
\be
\overline{H_{1N}^{(N-1)}} =  \langle \Phi_1|H^{(N-1)}|P \Psi_1^{(N)}\rangle  
\label{eq21} \  
\ee
and $\overline{H_{N1}^{(N-1)}}$ the transpose of the preceding matrix element.
By construction the Hamiltonian $H^{(N-1)}$ depends on the new coupling
constant $g^{(N-1)}$. Imposing Eq.~(\ref{eq16}) leads to a second order
equation for $g^{(N-1)}$. One gets explicitly
\be
{a^{(N-1)}{g^{(N-1)}}^2 + b^{(N-1)}g^{(N-1)} +  c^{(N-1)}} = 0
\label{eq22} \  
\ee
where 
\be
a^{(N-1)} = (H_{1N} - a_1^{(N)}H_{NN})F_{1N}
\label{eq23} \  
\ee
\be
b^{(N-1)} =  a_1^{(N)}(H_{NN}(\lambda_1^{(N)} - \epsilon_1) +  F_{1N}
(\lambda_1^{(N)} - \epsilon_N))
\label{eq24} \  
\ee
\be
c^{(N-1)} =   -a_1^{(N)}(\lambda_1^{(N)} - \epsilon_1)
(\lambda_1^{(N)} - \epsilon_N))
\label{eq25} \  
\ee
and
\be
F_{1N} = \sum_{i=1}^{N-1}a_i^{(N)}H_{Ni}
\label{eq26} \ . 
\ee
Since Eq.~(\ref{eq22}) is non-linear in $g^{(N-1)}$ and has two solutions, 
$g^{(N-1)}$ is chosen as the one closest to $g^{(N)}$ by continuity. 
The process can be iterated  step by step by projection from the space of 
dimension $N-1$ to
$N-2$ and further, generating subsequently a succession of coupling constants
$g^{(k)}$ at each iteration. At each step the projected wavefunction
$|P \Psi_1^{(k)}\rangle$ is obtained from $|\Psi_1^{(k)}\rangle$. Going over 
to the continuum limit if $N$ is large, one can derive a non-linear flow 
equation
\be
\frac{dg}{dx} = {\frac{1}{2a(x)g(x)+b(x)}}(\frac{dc}{dx} +
{\frac{db}{dx}}g(x) - {\frac{da}{dx}}{g(x)^2})
\label{eq27} \  
\ee
where $a(x), b(x), c(x)$ and $g(x)$ are the continuous extensions of the
corresponding discrete quantities. Eq.~(\ref{eq27}) is a
non-linear differential equation which a priori can only be solved numerically. 
It has to be mentioned here that the resolution of the discrete or continuous
flow equation requires the knowledge of the lowest eigenvalue and the
corresponding exact eigenfunction in the $k$ -
dimensional spaces. The eigenvalue may be experimentally known.   
The eigenvector components are not directly accessible in
practice. Information can be obtained from observables like the
quadrupole moment of the system in its ground state. They can be fixed
rigorously , f.i. by means of a Lanczos diagonalisation
procedure which determines the lowest eigenvalue and eigenvector. 

The present analysis shows how renormalisation concepts enter the problem 
of space reduction in the many-body problem.

{\it System at finite temperature T}.
We consider now the case of a quantum many-body system at temperature 
$T = \beta^{-1}$. If the dimension of the Hilbert space ${\cal H}^{(k)}$ is $k$ 
its canonical partition function reads
\be
Z^{(k)}(g_1^{(k)},g_2^{(k)},...,g_p^{(k)})= Tr_{k}e^{-\beta H^{(k)}}
\label{eq28} \  
\ee
if the Hamiltonian contains $p$ coupling constants. Going over to 
${\cal H}^{(k-1)}$ one imposes 
\be
Z^{(k)} = Z^{(k-1)}
\label{eq29} \ .
\ee
In the limit where  $\beta$ gets very large the trace reduces to the 
contribution of the lowest eigenvalue $\lambda_1^{(k)}$ ~\cite{avi} constrained
to stay constant through the dimensional reduction process induced by 
Eq.~(\ref{eq29}).

Using the Trotter formula~\cite{avi} 
\be
e^{-\beta H^{(k)}} \approx (1 - \frac{\beta H^{(k)}}{n})^{n}
\label{eq30} \  
\ee
for $n$ going to infinity and recalling Eq.~(\ref{eq12})
the trace defined in Eq.~(\ref{eq28}) gets a polynomial of order $n$ in
$g^{(k)}$ 
\be
Z^{(k)} =  \sum_{i=0}^{n}h_i^{(k)} {g^{(k)}}^i 
\label{eq31} \  
\ee
where $\left\{ h_i^{(k)}, \,i=1,\cdots, n\right\}$ are fixed real coefficients.
Eq.~(\ref{eq29}) leads to  
\be
\sum_{i=0}^{n}h_i^{(k)}{g^{(k)}}^i = \sum_{i=0}^{n}h_i^{(k-1)}{g^{(k-1)}}^i
\label{eq32} \ .
\ee
Going to the continuum limit $k \mapsto x$ and taking the derivative of the
partition function leads to the flow equation
\be
\frac{dg(x)}{dx} = - 
\frac{\sum_{i=0}^{n} \frac{dh_i(x)}{dx}g^i(x)}{\sum_{i=0}^{n} i h_i(x) 
g^{i-1}(x)} 
\label{eq33} \ .
\ee 
Since $n$ has to be very large in practice one may go over to an integral
formulation of the discrete sums.  

{\it Exceptional points and fixed points}. 
As a last interesting point we finally address the problem concerning the
divergence of perturbation expansions and the existence of fixed points of the 
running coupling constant $g(x)$. 

It has been rigorously established that the eigenvalues $\lambda_k{(g)}$
of $H(g) = H_0 + gH_1$ are analytic functions of $g$ with only algebraic
singularities~\cite{kat}. They get singular at so called exceptional points 
$g = g_e$ which are first order branch points in the complex $g$ - plane. 
Branch points appear if two (or more) eigenvalues get degenerate. This can 
happen if $g$ can take values such that $H_{kk} =  H_{ll}$
which corresponds to a so-called level crossing. As a consequence, if a level
belonging to the $P{\cal H}$ subspace defined above crosses a level
lying in the complementary $Q{\cal H}$ subspace the perturbation development
constructed from $H_{eff}(E)$ diverges~\cite{sch2}. Exceptional points are 
defined as the solutions of 
\be
f(\lambda(g_e))  = det[ H(g_e) - \lambda(g_e)I] = 0
\label{eq34} \ 
\ee
and 
\be
\frac{df(\lambda(g_e))}{d\lambda}|_{\lambda= \lambda(g_e)}= 0
\label{eq35} \ 
\ee
where $f(\lambda(g))$ is the secular determinant.
It is now possible to show that exceptional points are
connected to fixed points corresponding to $dg/dx = 0$ in specific cases. If   
$\left\{\lambda_i(g)\right\}$ are the set of eigenvalues the secular equation
can be written as
\be
\prod_{i=1}^N {(\lambda - \lambda_i)} = 0
\label{eq36} \ .
\ee
Consider $\lambda = \lambda_p$ which satisfies Eq.~(\ref{eq34}). Then
Eq.~(\ref{eq35}) can only be satisfied if there exists another eigenvalue
$\lambda_q = \lambda_ p$, hence if the spectrum is degenerate. This is the case
at an exceptional point. 
Going back to the algorithm described above consider the case where the 
eigenvalue $\lambda_1$ gets degenerate with some other eigenvalue 
$\lambda_i^{(k)}({g = g_e})$ at some step $k$ in the space reduction process. 
Since $\lambda_1$ is constrained to be constant,
\be
\lambda_i^{(k)}({g_e}) = \lambda_i^{(l)}({g'_e})
\label{eq37} \ 
\ee
which is realised in any projected subspace of size $k$ and $l$ containing  
states $|\Phi_1\rangle$ and $|\Phi_i\rangle$. Going over to the continuum 
limit and considering the subspaces of dimension $x$ and $x + dx$ one can write 
\be
\frac{d\lambda_1}{dx} = 0 = \frac{d\lambda_i(x)}{dx}
\label{eq38} \ .
\ee
Consequently
\be
\frac{d\lambda_i}{dg_e} \frac{dg_e}{dx} = 0
\label{eq39} \ .
\ee
Due to the Wigner - Neumann avoided crossing rule the degeneracy of eigenvalues 
is generally not fulfilled  for real values of the coupling constant. There 
exist however specific situations, like systems with special symmetry 
properties ~\cite{sch2,he,yu} or infinite systems~\cite{sach1} for which 
degeneracy can occur. In these cases Eq.~(\ref{eq39}) is realised if
\be
\frac{dg_e}{dx} = 0\,\,\, \quad \mathrm{and}\,\,\,\, 
\frac{d\lambda_i}{dg_e}\not = 0
\label{eq40} \ 
\ee  
The second relation works if crossing takes place 
and $g_e$ is a fixed point in the sense of renormalisation theory.

Eq.~(\ref{eq40}) shows the connection between exceptional and fixed points 
in the framework
of the present approach. Ground state degeneracy is indeed a signature for the
existence of phase transitions~\cite{sml}, perturbation expansions break down
at transition points. In fact, one identifies in the present result the
properties of a quantum phase transition which is induced by level crossing
~\cite{sach1,sach2}. The ground state wavefunction changes its properties when 
the (real) coupling constant $g$ crosses the exceptional point $g_e$. There
the eigenstates exchange the main components of their projection on the set of 
basis states ${|\Phi_i \rangle, i = 1,...N}$.

{\it Conclusions}
In summary, we developed a non-perturbative effective theory of the bound
state many-body quantum problem based on a reduction process of the dimensions
of the initial Hilbert space. The central point concerns the renormalisation of
the coupling constant(s) which characterise(s) the initial Hamiltonian under
the constraint that the lowest eigenenergy is known. We presented  
a projection approach for the case of a system at temperature $T=0$ and
constructed the flow equation for the coupling constant. We worked out the case
of a system at finite temperature. Finally we showed the 
relationship between exceptional points
corresponding to level crossings in the spectrum where perturbation expansions
break down and fixed points of the coupling constants which characterise phase
transitions. 

The present approach can be straightforwardly extended to the case of an 
Hamiltonian characterised by several coupling constants. Effective expressions
of observables other than the energy and flow equations for the physical
constants which characterise them can be worked out.  

I acknowledge interesting and fruitful discussions with J. Polonyi and A.
Kenoufi which revived my interest in this subject and inspired the present work.
I gratefully thank J. M. Carmona and M. Henkel for their careful reading of 
the manuscript, comments and advices.

\end{document}